# Electroluminescence from a diamond device with ion-beam-micromachined buried graphitic electrodes

J. Forneris[a,b,c#], A. Battiato[a,b,c], D. Gatto Monticone[a,b,c], F. Picollo[b,c,a]

G. Amato[d], L. Boarino[d], G. Brida[d], I. P. Degiovanni[d], E. Enrico[d], M. Genovese[d],
E. Moreva[d], P. Traina[d], C. Verona[e], G. Verona-Rinati[e], P. Olivero[a,b,c]

a Physics Department and NIS Interdepartmental Centre, University of Torino, Torino, Italy
b Istituto Nazionale di Fisica Nucleare (INFN), Sez. Torino, Torino, Italy
c Consorzio Nazionale Interuniversitario per le Scienze Fisiche della Materia (CNISM), Sez. Torino, Torino, Italy
dad Istituto Nazionale di Ricerca Metrologica (INRiM), Torino, Italy
e Department of Industrial Engineering, University of Roma "Tor Vergata", Roma, Italy

**Abstract**
Focused MeV ion microbeams are suitable tools for the direct writing of conductive graphitic channels buried in an insulating diamond bulk, as demonstrated in previous works with the fabrication of multi-electrode ionizing radiation detectors and cellular biosensors. In this work we investigate the suitability of the fabrication method for the electrical excitation of colour centres in diamond. Differently from photoluminescence, electroluminescence requires an electrical current flowing through the diamond sub-gap states for the excitation of the colour centres. With this purpose, buried graphitic electrodes with a spacing of 10 μm were fabricated in the bulk of a detector-grade CVD single-crystal diamond sample using a scanning 1.8 MeV He$^+$ micro-beam. The current flowing in the gap region between the electrodes upon the application of a 250 V bias voltage was exploited as the excitation pump for the electroluminescence of different types of colour centres localized in the above-mentioned gap. The bright light emission was spatially mapped using a confocal optical microscopy setup. The spectral analysis of electroluminescence revealed the emission from neutrally-charged nitrogen-vacancy centres (NV$^0$, $\lambda_{ZPL}$ = 575 nm), as well as from cluster crystal dislocations (A-band, $\lambda$ = 400-500 nm). Moreover, an electroluminescence signal with appealing spectral features (sharp emission at room temperature, low phonon sidebands) from He-related defects was detected ($\lambda_{ZPL}$ = 536.3 nm, $\lambda_{ZPL}$ = 560.5 nm); a low and broad peak around $\lambda$ = 740 nm was also observed and tentatively ascribed to Si-V or GR1 centres. These results pose interesting future perspectives for the fabrication of electrically-stimulated single-photon emitters in diamond for applications in quantum optics and quantum cryptography.

## 1. Introduction
Colour centres in diamond are attracting an ever-growing interest due to their appealing photon-emission properties (high stability and quantum efficiency at room temperature) for applications in quantum optics and photonics [1].

The development of diamond devices based on the functionalities of sub-superficial graphitic micro-electrodes has been investigated in several recent works [2, 3]. The fabrication method relies on the direct writing in the single-crystal diamond bulk of amorphized channels by the selective dam-age induction associated with the Bragg's peak of MeV ions. The crystal volume in which the radiation-induced vacancy density exceeds a threshold value converts to nano-crystalline graphite upon thermal treatment at temperatures above 900 °C, while the diamond lattice is partially recovered where the radiation damage density is lower [4, 5].

---

[#] Correspondig author; Jacopo Forneris, Physics Department, University of Torino - via P. Giuria, 1 - 10126 Torino, Italy; e-mail: jacopo.forneris@unito.it

Since the radiation damage from MeV ions is prominently induced at the ion end of range, the fabrication method has been proved to be effective for the fabrication of conductive graphitic electrodes located several micrometers below the surface of the electrically insulating and optically transparent diamond dielectric. The fabrication process is characterized by a spatial resolution which is limited only by the ion beam spot size and by the ion lateral straggling in the material. The technique has been effectively exploited to fabricate and characterize diamond-based devices such as ionizing radiation detectors [3, 6], cellular bio-sensors [7] and IR emitters [8].

In this work, we investigate the potential exploitation of the fabrication technique for the electrical excitation of colour centres in diamond, aiming at the development of practical single-photon-emitting devices based on the efficiency and stability of the above-mentioned centres, for applications in quantum optics [9, 10].

With this purpose, a diamond device with four independent buried graphitic electrodes was fabricated by means of MeV ion beam lithography. Differently from previous works, in which the electrical excitation of colour centres is achieved by means of diodes and p-i-n structures [9-11], the device proposed in this work exploits the current flowing between buried graphitic electrodes, in order to obtain electroluminescence (EL) from the diamond region located in the inter-electrode gap. The light emission properties of the fabricated structure were investigated by means of luminescence mapping and the relevant spectra were analyzed to attribute the active defects in the device under investigation.

## 2. Experimental

*Sample preparation*
The device under test, whose IBIC characterization was previously reported in [6], is based on a ~40 μm thick intrinsic diamond film homoepitaxially grown on a commercially available 4×4×0.4 mm$^3$ type-Ib single-crystal high-pressure high-temperature (HPHT) substrate, using a microwave plasma enhanced chemical vapor deposition process at the laboratories of Rome "Tor Vergata" University. A scanning 1.8 MeV He$^+$ ion microbeam (~10 μm spot size) was used to perform a deep ion beam lithography (DIBL) process [2, 6] at the AN2000 micro-beam line of the INFN National Laboratories of Legnaro (I). Five graphitic channels were written at a depth of ~3 μm below the diamond surface using a ion fluence of ~1.5×10$^{17}$ cm$^{-2}$, which, according to SRIM2011 simulations [12], is sufficient to achieve a vacancy density above the graphitization threshold at the end of range (Fig. 1a) [6].

Before the DIBL process, a slowly thinning copper mask was deposited on the diamond surface in order to control the depth of the 1.8 MeV He$^+$ Bragg's peak in diamond, ensuring the emersion of the buried graphitic channels at their endpoints, as described in previous works [2, 3, 6].

The implanted sample underwent a thermal annealing process for 2 hours in vacuum, in order to promote the conversion of the amorphized regions at the end of the Bragg's peak to a graphitic phase and to concurrently recover the residual structural damage in the region comprised between the buried channels and the diamond surface. The resulting micro-fabricated structure (Fig. 1b) consisted of four parallel, ~10 μm wide independent graphitic electrodes, spaced by ~12 μm, plus an additional horizontal electrode, which was not employed in the present work. Each graphitic channel was connected to the external circuitry by the deposition of 80 nm thick Cr/Al circular contacts (150 μm diameter), which were exploited as pads for the electrodes wire-bonding. As schematically represented in Fig. 1b, the graphitic channels were shorted in pairs, defining a two-electrode inter-digitated geometry.

An electrical characterization of the fabricated structure was performed by means of current-voltage measurements (not reported here) between the two independent electrodes. The measured current was lower than 50 pA at ±100 V applied bias, thus confirming the mutual insulation of the electrodes through the annealed diamond matrix and indicating a negligible surface leakage current. On the other hand, at applied voltages larger than ±200 V an abrupt current increase to few hundreds of nA was observed, and the current reached values of tens of μA at ±500 V. Such phenomenon was repeatable and non-destructive for the sample under test. Similar observations were interpreted in a previous work on ion implanted diamond as the occurring of an avalanche breakdown [5], although further investigation will be necessary to unequivocally interpret this charge conduction mechanism.

*Optical characterization*
The electric current flowing between the buried channels was exploited to provide the electrical excitation of the colour centres in the inter-electrode gap. A bright EL emission, visible to the bare eye, appeared at bias voltages above 200 V (typical currents ~100 nA) and increased in intensity at increasing bias. In Fig. 2, an optical micrograph of the inter-electrode region of the sample taken at a bias voltage of 250 V is shown, where EL is clearly visible and the presence of different spectral components of the emitted light is apparent.

In order to clarify the emission properties of the electroluminescent device and to identify with higher spatial resolution the light-emitting regions of the structure, an EL mapping of the sample was performed by adapting a home-built confocal photoluminescence microscopy setup (Fig. 3a) [13]. The sample was mounted on a remotelycontrolled three-axis piezo-electric stage, with a scan area of 100×100 μm$^2$ and a nanometer positioning accuracy, and was raster-scanned to image the EL properties of the structure under test. The induced luminescence was collected by a 100× air objective (numerical aperture N.A. = 0.9) and then focused with an achromatic doublet into a graded-index multimode optical fibre, which both provided an optical connection to the detection system and acted as the pinhole aperture for the confocal system. The detection system consisted of a photon-counter based on one Si-single-photon-avalanche photo-diode (SPAD) Perkin-Elmer SPCAR-Q14-FC operating in Geiger mode; a digital counter recorded the total counts detected by the SPAD, enabling the measurement of the total EL intensity for each pixel.

## 3. Results

An EL map acquired at a bias voltage of 250 V is shown in Fig. 3b. The encoded colour scale displays for clarity a maximum emission rate value of 5×10$^4$ counts per second (cps), each count corresponding to the detection of a single photon; however, the values recorded at the centre of the bright spot reached values up to 10$^6$ cps. The map clearly shows that the electroluminescent region corresponds to a localized conduction path connecting the buried electrodes (highlighted by the dashed black lines in Fig. 3b). The conductive path corresponds to the bright region highlighted in the optical micrograph in Fig. 2 and originates at the endpoint of the grounded electrode, where the sharpening of the buried graphitic channel increases the local electric field as an effect of the tip geometry.

With the purpose of performing spectroscopic measurements, the radiation from the light-emitting region highlighted by the green circle in Fig. 2, comprising most of the EL signal was collected with a Jobin Yvon Raman micro-spectrometer equipped with a CCD Andor "DU420A-OE" detector. As the light emission is not stimulated by a localized standard laser excitation, but instead is induced by the electrical current flowing between the electrodes, the collected light is not spatially limited to the focal point of the experimental setup and the spatial resolution for the spectral measurement was estimated to be of ~10 μm. A typical EL spectrum is shown in Fig. 4. The two dips at 633 nm and at 428 nm are instrumental artifacts caused by the presence of a notch filter mounted in the spectrometer, which attenuates the laser excitation when the setup operates in its usual configuration. Starting from lower wavelengths, different features can be identified in the EL spectrum. Firstly, a ~100 nm broad peak centred at 435 nm is visible. The broad peak usually referred as the A-band and its origin is commonly attributed to radiative carriers recombination at extended lattice dislocations. Evidence of the A-band was found in previous EL experiments in diamond P-I-N junctions [11, 14, 15]. Therefore, the observation of the 435 nm broad peak proves that the injected charge carriers are interacting with electronic states within the bandgap [14]. Secondly, the two sharp peaks at 536.3 nm and 560.5 nm are attributed to a helium-related defect previously observed in photoluminescence and cathodoluminescence [5, 15], consistently with the fact that He$^+$ was employed for the micro-fabrication process and stray ions could easily have been implanted in the inter-electrode gap region. Thirdly, the peak at 575 nm and its replica at higher wavelengths respectively correspond to the zero-phonon-line and phonon sidebands of the well-known NV$^0$ centre, as extensively observed in cathodoluminescence and EL measurements [5, 9, 10]. Finally, a small peak emerging from the background at ~738-740 nm can be tentatively attributed to the Si-V colour centres [15], since Si could be present as a contaminant in the CVD chamber, and it has been reported to be active in EL in previous works [16]. Alternatively, the peak could be ascribed to the general radiation centre GR1 (741 nm), which is associated with radiation-induced vacancies. Although a direct observation in electroluminescence has not been reported so far to our knowledge, the emission from GR1 has been observed in cathodoluminescence regime [15] where a similar excitation process through the injection of electrodes in the diamond dielectric is taking place. Finally, it is worth noting the absence of the radiation B-band (770 nm) associated with radiation damage and reported in EL works [10]. Despite the spatial resolution of the experimental setup was not sufficient to provide a position-dependent analysis, the micrograph in Fig. 2 shows that the contributions of the A-band and of the He-related centre vary from region to region: areas with a greater contribution from the A-band are clearly identified by the blue light, while regions dominated by the emission at longer wavelength, where no contribution is associated with extended lattice dislocations, appear as yellow.

## 4. Conclusions

A structure consisting of parallel buried graphitic electrodes in single-crystal diamond was fabricated with DIBL with a 1.8 MeV He$^+$ scanning microbeam. The current flowing in the gap region between the buried electrodes has been exploited to stimulate EL from colour centres located in the above-mentioned region. EL mapping highlighted that the emission occurs along localized paths between the electrodes. The EL spectrum exhibited light emission from dislocation clusters (A-band), $NV^0$ centres and helium-related defects associated with the ion implantation. The results presented in this work confirm that it is possible to electrically excite colour centres by means of charge injection from graphitic non-rectifying electrodes.

Furthermore, the results highlight the possibility of electrically exciting He-related defects, which are characterized by appealing photophysical properties, i.e. high emission rates, sharp spectral emission, negligible phononic sidebands. The possibility to identify and electrically stimulate isolated He-related defects might open to the study of new electroluminescent single-photon sources in diamond for applications in photonics and quantum information.

Acknowledgements

This work is supported by the following projects: FIRB "Futuro in Ricerca 2010" (CUP code: D11J11000450001) funded by the Italian Ministry for Teaching, University and Research (MIUR); "A.Di.N-Tech." project (CUP code: D15E13000130003) funded by the University of Torino and Compagnia di San Paolo in the framework of the 'Progetti di ricerca di Ateneo 2012' scheme; EMRP project 'EXL02-SIQUTE', jointly funded by the EMRP participating countries within EURAMET and the European Union.

**Figures and captions**

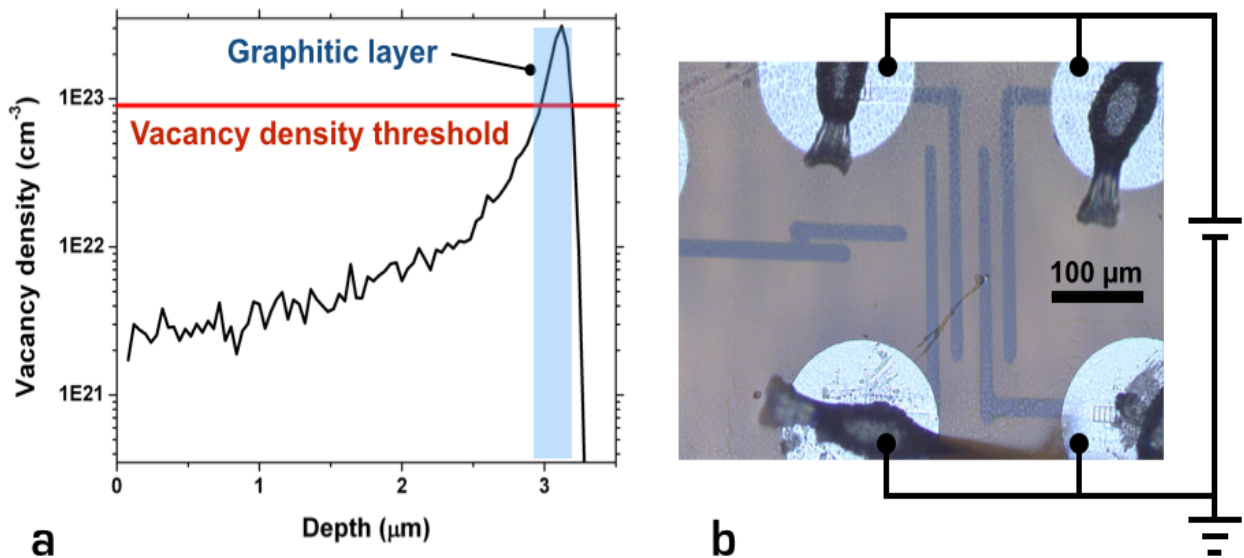

Fig. 1: a) Simulated vacancy density profile of 1.8 MeV He$^+$ in diamond at a fluence of 1.5×10$^{17}$ cm$^{-2}$. The vacancy density exceeds the graphitization threshold at 3 μm below the sample surface. b) Optical micrograph of the graphitic electrodes structure after annealing and wire bonding. The electrical connections are schematically presented.

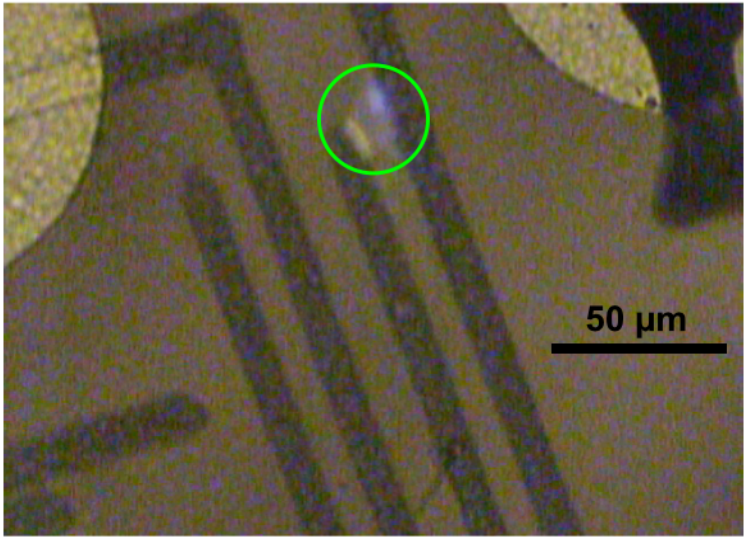

Fig. 2: Optical micrograph of the sample biased at a voltage of 250 V. Blue and yellow electroluminescence are clearly visible in the circled area. The green circle indicates the region from which the EL spectrum shown in Fig. 4 was acquired.

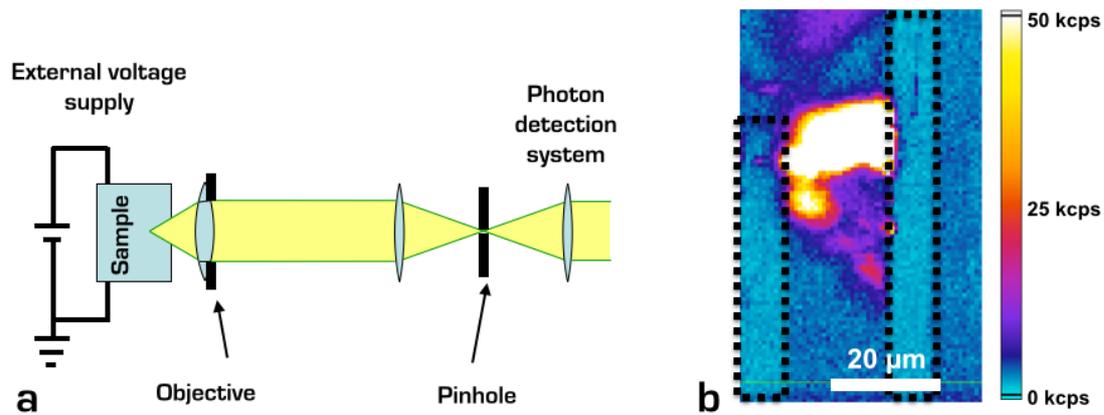

Fig. 3: a) Schematic representation of the confocal microscopy setup adopted for electroluminescence (EL) mapping. The sample is mounted on a xyz remotely-controlled piezo-electric stage and is biased by an external voltage source. Emitted light is collected by a 100× objective and is focused through a multi-modal optical fiber, which is acting as a pinhole. Such a configuration allows for the optical sectioning of the device. b) EL map acquired from the device at an applied voltage of 250 V. The dashed black lines identify the relative position of the graphitic electrodes.

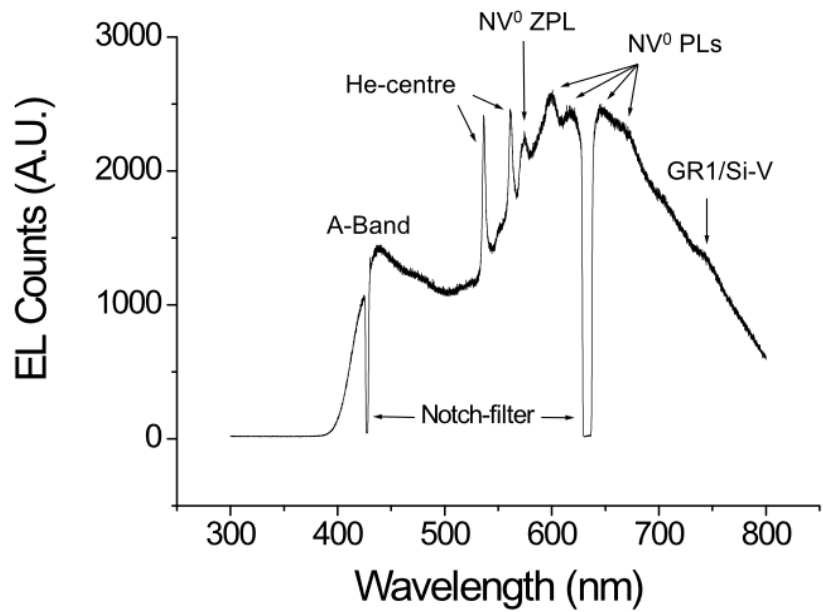

Fig. 4: EL spectrum from the yellow region indicated by the red arrow in Fig. 2. The following features are highlighted: A-band; He-centre; $NV^0$ and (tentatively) GR1 or Si-V. A notch filter present in the spectrometer creates two dips at 428 and 633 nm.